\documentstyle[a4,12pt]{article}
\begin{document}
\begin{flushright}
NYU-TH 00/10/07 \\
hep-th/0010186 \\
\end{flushright}
\vskip 1cm
\begin{center}
{\Large \bf Cosmology on a Brane in Minkowski Bulk}\\
\vskip 2cm
{C\'edric Deffayet}\footnote{e-mail cjd2@physics.nyu.edu}\\ 
\vskip 1cm 
{\it Department of Physics, New York University, New York, NY 10003.}\\
\vskip 1.6cm
 {\bf Abstract.}  
\vskip 1cm
\end{center}

{We discuss the cosmology of a 3-brane embedded in a 5D bulk space-time with a cosmological constant
when an intrinsic
curvature Ricci scalar is included in the brane action. After deriving the `brane-Friedmann' equations
for a $Z_2$ symmetrical metric, 
we focus on the case of a Minkowski bulk.
We show that there exist two classes of solutions, close to the usual 
Friedmann-Lema\^{\i}tre-Robertson-Walker
cosmology for small enough Hubble radii. When the Hubble radius gets larger one either 
has a transition to a fully 5D regime or
to a self-inflationary solution  which produces a late accelerated expansion. 
We also compare our results with 
a perturbative approach and eventually discuss
the embedding of the brane into the Minkowski space-time.
This latter
 part of our
discussion also applies when no intrinsic curvature term is included.}

\def\beq{\begin{equation}}
\def\eeq{\end{equation}}
\def\d{\delta}
\def\fourG{{{}^{(4)}G}}
\def\4R{{{}^{(4)}R}}
\def\H{{\cal H}}
\def\K5{{\kappa}}
\def\K52{{\kappa^2}}
\def\C{{\cal C}}
\def\lamb{{\rho_{\Lambda}}}

\newcommand{\A}{A}
\newcommand{\B}{B}
\newcommand{\mmu}{\mu}
\newcommand{\mnu}{\nu}
\newcommand{\ii}{i}
\newcommand{\jj}{j}
\newcommand{\jl}{[}
\newcommand{\jr}{]}
\newcommand{\ml}{\sharp}
\newcommand{\mr}{\sharp}

\newcommand{\da}{\dot{a}}
\newcommand{\db}{\dot{b}}
\newcommand{\dn}{\dot{n}}
\newcommand{\dda}{\ddot{a}}
\newcommand{\ddb}{\ddot{b}}
\newcommand{\ddn}{\ddot{n}}
\newcommand{\pa}{a^{\prime}}
\newcommand{\pb}{b^{\prime}}
\newcommand{\pn}{n^{\prime}}
\newcommand{\ppa}{a^{\prime \prime}}
\newcommand{\ppb}{b^{\prime \prime}}
\newcommand{\ppn}{n^{\prime \prime}}
\newcommand{\fda}{\frac{\da}{a}}
\newcommand{\fdb}{\frac{\db}{b}}
\newcommand{\fdn}{\frac{\dn}{n}}
\newcommand{\fdda}{\frac{\dda}{a}}
\newcommand{\fddb}{\frac{\ddb}{b}}
\newcommand{\fddn}{\frac{\ddn}{n}}
\newcommand{\fpa}{\frac{\pa}{a}}
\newcommand{\fpb}{\frac{\pb}{b}}
\newcommand{\fpn}{\frac{\pn}{n}}
\newcommand{\fppa}{\frac{\ppa}{a}}
\newcommand{\fppb}{\frac{\ppb}{b}}
\newcommand{\fppn}{\frac{\ppn}{n}}

\newcommand{\dA}{\dot{A_0}}
\newcommand{\dB}{\dot{B_0}}
\newcommand{\fdA}{\frac{\dA}{A_0}}
\newcommand{\fdB}{\frac{\dB}{B_0}}

\section{Introduction}
A lot of interest has recently been raised for field theories where the standard model
of high energy physics is assumed to live on a surface (called generically a
brane) embedded in a larger space-time.
The (super)gravitational fields are in contrast usually considered
 to live in the whole space-time. 
Models coming from string-M theory like the Ho\v{r}ava-Witten walls \cite{hw96}, or D-branes,
 as well as from a more phenomenological approach 
\cite{add98,rs99a}  have been extensively studied
in particular in cosmology. 

A question, which naturally arises, 
is how to recover standard gravity
in its well tested perturbative regime. A first approach is to assume that the
dimensions transverse to the brane are compact, in which case the usual Kaluza-Klein results
allow to recover 4D gravity on scales which are larger than the size of the
extra dimensions. The main difference with the old picture being there that,
because the standard model fields are assumed to be brane-localized, one can
 have very large extra-dimensions in comparison to the length scales
probed by high energy physics \cite{add98}. 
On scales smaller than the size of the extra dimensions, on the other hand,  gravity enters a
higher dimensional regime. Another way of recovering usual 4D gravity on the brane for large distances is to embed 
a positive tension 3-brane into an $AdS_5$ bulk \cite{rs99a}
in which case the crossover scale between 4D and 5D gravity is set by the $AdS$ radius. In this
latter case, the extra dimension  has also a finite size.

We will be mainly here interested in a more recent approach advocated by Dvali {\it et al.} 
\cite{Dvali:2000hr,Dvali:2000xg}. In this approach, the 3-brane is 
embedded in a space-time with infinite size extra dimensions,
with the hope that this picture could shed new light on the standing problem of
the cosmological constant as well as on supersymmetry breaking \cite{Dvali:2000hr,Witten:2000zk}.
 The recovery of the usual gravitational laws is obtained by adding to the action of the
brane an Einstein-Hilbert term computed with the brane intrinsic curvature. 
The presence of such a term in the action is generically induced by quantum corrections coming from the bulk
gravity and its coupling with matter living on the brane and should be included for a large class of theories for 
 self-consistency (see 
e.g. \cite{Dvali:2000xg,oldgrav}). 
In the particular
case of a 3-brane embedded in a 5D Minkowski space-time, Dvali, Gabadadze and Porrati
 have  shown that one recovers a standard 4D Newtonian potential for small distances, 
 whereas gravity is in a 
5D regime for large distances. The tensorial structure of the graviton propagator 
in this theory has in contrast been shown
to be higher dimensional which is likely to rule out the theory from a phenomenological point of view. 
For a brane embedded in a bulk with 2 or more extra dimensions, one can show that the
theory is always 4-dimensional \cite{Dvali:2000xg} for a zero thickness brane
\footnote{however, a simple regularization scheme in the graviton propagator accounting for a brane non zero thickness
 leads to a behaviour similar to the 5D case  \cite{massimo}.\label{noteenbas}}.

Our purpose is here to study the cosmology of these models in the case of a 5D bulk.
Although, as mentionned above, such a theory has serious phenomenological problems, its cosmology can help to have a
better understanding of this kind of model and of the idea of gravity localization through an intrinsic curvature term
on the brane. In particular, exact cosmological solutions, as given in this work, provide a unique way
to test the theory in its fully non linear regime.
The Friedmann-like equations governing the cosmological evolution of a brane possessing an
intrinsic curvature term in its action have already been derived and discussed for 
an AdS-Schwarzschild bulk space-time \cite{Collins:2000yb,Shtanov:2000vr,Nojiri:2000gv}. In the first part of this paper we
derive similar equations (valid whenever the bulk matter is a pure cosmological constant) 
in a slightly different way, following the work of Bin{\'e}truy {\it et al.} \cite{Binetruy:2000ut,bdel}.
For this pupose we will adopt a brane-based coordinate system and specialize to a $Z_2$ symmetrical metric.
We then discuss these equations for a vanishing bulk cosmological constant as well as for a vanishing Schwarzschild
mass parameter.
We show that there exist two possible types of cosmology which are both similar to the usual 4D 
Friedmann-Lema\^{\i}tre-Robertson-Walker (FLRW) cosmology
 when the Hubble radius is small.
We point out, however, a discrepancy between the Newton's constant inferred by cosmology in that regime, and the one {\it defined} by
a Cavendish-like experiment. 
 For larger Hubble radius, the cosmology either 
 evolves  to the
brane-typical linear relationship between the Hubble parameter and energy density, or to a brane
self-inflationary solution previously noticed by Shtanov \cite{Shtanov:2000vr}. In this latter case, one
has a late-time accelerated expansion sourced by the intrinsic curvature term   of the brane itself
(and not by its tension). 
In the last part of this paper we interpret the found two-folded cosmology
by looking  at the brane embedding into the bulk space-time.
We give in particular the change of coordinates between our brane-based metric and a canonical Minkowskian bulk metric.

\section{4D gravity on a 3-brane in 5D Minkowski bulk}
We consider a 3-brane embedded in a 5D space-time with an intrinsic curvature term included in the brane action.
We can take accordingly the following action 
\beq 
S_{(5)}=-\frac{1}{2\kappa^2} \int d^5X
\sqrt{-\tilde{g}} \tilde{R} 
+\int d^5X \sqrt{-\tilde{g}}{\cal L}_m,
\label{action} 
\eeq 
to which we add the brane curvature term
\beq
-\frac{1}{2\mu^2} \int d^4x
\sqrt{-{g}} {R}. \label{branac}
\eeq
The first term in (\ref{action}) 
corresponds to the Einstein-Hilbert
action in five  dimensions for a five-dimensional metric ${\tilde g}_{AB}$ (bulk metric)
 of  Ricci scalar
$\tilde{R}$. Similarly, (\ref{branac}) is the Einstein-Hilbert action for the induced metric $g_{cd}$ on the brane, $R$ being its scalar
curvature. The induced metric\footnote{Throughout this article, we will adopt the following  convention for indices:
upper case Latin letters $A,B,...$ will denote 5D indices: $0,1,2,3,5$; lower case Latin letters from 
the beginning of
the alphabet: $c,d,...,$ will denote 4D indices parallel to the brane: $0,1,2,3$;
lower case Latin letters from the middle of the
alphabet: $i,j,...,$ will denote space-like 3D indices parallel to the brane: $1,2,3$.}
 $g_{cd}$ is defined as usual from the bulk metric
${\tilde g}_{AB}$ by 
\beq \label{induite}
g_{cd} = \partial_c X^A \partial_d X^B {\tilde g}_{AB},
\eeq
where $X^A(x^c)$ represents the coordinates of an event on the brane labelled by $x^c$. 
We point out that the Einstein-Hilbert term (\ref{branac}), if not present at classical level, 
should appear quite generically (namely for any brane with a non conformal world volume theory \cite{Dvali:2000xg}) from
quantum corrections so that its inclusion is dictated by self consistency.
The second term in (\ref{action}) corresponds to the `matter' content. Aside from bulk matter, 
we have included there the contribution of the brane-localized matter. This latter contribution 
 can be further rewritten 
\beq \label{lagofbra}
\int d^4x \sqrt{-{g}} \left(\lambda_{brane} + l_m\right),
\eeq
where $l_m$ represents the lagrangian density of `matter' fields living on the brane, and $\lambda_{brane}$ 
the brane tension (or `cosmological constant'). 
We stress here that this tension is unrelated to the  presence of the 
intrinsic curvature term (\ref{branac}), and can in principle be tuned to be zero \cite{Dvali:2000hr,Dvali:2000xg}.
We also define from the dimensionful constants $\kappa$ and $\mu$ the related quantities
\beq
\kappa^2 = 8 \pi G_{(5)} = {M}_{(5)}^{-3},
\eeq
and  
\beq
\mu^2 = 8 \pi G_{(4)} = {M}_{(4)}^{-2}. \label{geengri}
\eeq

For a brane embedded in a Minkowski space-time with an action obtained by adding (\ref{action}) 
and (\ref{branac}),
Dvali {\it et al.} \cite{Dvali:2000hr} have shown that the usual 4D Newton's law, 
for static point like sources on the brane, 
is recovered at
small distances on the brane; whereas at large distances the gravitational force is given by the 5D $1/r^3$ law. 
The crossover length scale between the two different regimes is given by 
\beq \label{crosso}
r_0 =\frac{M_{(4)}^2}{2M_{(5)}^3}. 
\eeq
However,  the model is plagued by the presence of an extra scalar polarization degree of freedom which 
can be seen by looking at the tensor structure of the graviton propagator. This structure (analogous to the one of a 5D
massless graviton) is also responsible for a rescaling of the Newton's constant with respect to the usual 4D one. Namely,
if one wants to define the Newton's constant through some Cavendish like experiment, 
looking at the force exerted between static
point-like particles in the regime where this force obey the usual 4D Newton's law (i.e. for distances smaller than
$r_0$), then one finds the following relationship \cite{Dvali:2000rv} between the so-defined 
Newton's constant $G_N$ and $G_{(4)}$:
\beq \label{newton}
G_{N} = \frac{4}{3} \frac{\mu^2}{8 \pi} = \frac{4}{3} G_{(4)}.
\eeq
This rescaling is due to the presence of the extra scalar degree of freedom which
 exerts an extra attraction with
respect the the ordinary case, it does not persist for more than one extra dimension and a zero
 thickness brane \cite{Dvali:2000xg}.

\section{Brane Friedmann equations} \label{s3}
The purpose of this section is to derive Friedmann like equations for the brane metric. 
We will consider  
five-dimensional spacetime metrics of the form 
\begin{equation} \label{metmom}
ds^{2}=\tilde{g}_{AB}dx^A dx^B = 
g_{cd} dx^{c}dx^{d} + b^{2}dy^{2}, 
\end{equation}
where $y$ is the coordinate of the fifth dimension and we will adopt a brane-based approach where the brane is the hypersurface
defined by $y=0$. Being interested in
 cosmological solutions, we take a metric of the form
\begin{equation} \label{cosmback}
ds^{2}=-n^{2}(\tau,y) d\tau^{2}+a^{2}(\tau,y)\gamma_{ij}dx^{i}dx^{j}+b^{2}(\tau,y)dy^{2},
\label{metric}
\end{equation}
where $\gamma_{ij}$ is a maximally symmetric 3-dimensional metric ($k=-1,0,1$ will 
parametrize the spatial curvature).

The five-dimensional Einstein equations take the form 
  \beq
{\tilde G}_{\A\B}\equiv{\tilde R}_{\A\B}-{1\over 2}{\tilde R}{\tilde g}_{\A\B}
  =\kappa^2 \tilde{S}_{\A\B}, \label{einstein}
\eeq
where ${\tilde R}_{\A\B}$ is the five-dimensional Ricci tensor,
and the tensor $\tilde{S}$ is the sum of the energy momentum tensor $\tilde{T}$ of matter 
and the contribution coming from the scalar curvature of the
brane. We denote this latter contribution $\tilde{U}$. 
We have
\begin{equation} \label{sourcier}
\tilde{S}^\A_{\quad \B} = \tilde{T}^\A_{\quad \B} + \tilde{U}^\A_{\quad \B},
\end{equation}
where the energy momentum tensor can  be further decomposed 
 into two parts 
\begin{equation}
\tilde{T}^\A_{\quad \B}  =  \check{T}^\A_{\quad \B}|_{_{\rm bulk}} 
+ T^\A_{\quad \B}|_{_{\rm brane}}. \label{NRJ}
\end{equation}
The first tensor $\check{T}^\A_{\quad \B}|_{_{\rm bulk}}$ 
is the energy momentum tensor 
of the bulk matter, which will be assumed in the present work to be that of  
a cosmological constant: 
\begin{equation}
\check{T}^\A_{\quad \B}|_{_{\rm bulk}}= \mbox{diag} 
\left(-\rho_B,-\rho_B,-\rho_B,-\rho_B,-\rho_B \right), 
\label{bulksour}
\end{equation}
where the energy density $\rho_B$ is a constant.
The second 
term $T^\A_{\quad \B}|_{_{\rm brane}}$ in (\ref{NRJ}) corresponds to the matter content
in the 
brane $(y=0)$. Since we consider here only strictly homogeneous and isotropic 
geometries inside the brane, 
the latter can be expressed quite generally in the form
\begin{equation}
T^\A_{\quad \B}|_{_{\rm brane}}= \frac{\delta (y)}{b} \mbox{diag} 
\left(-\rho_b,p_b,p_b,p_b,0 \right), 
\label{source}
\end{equation}
where the energy density $\rho_b$ and pressure $p_b$ are independent of the 
position inside the brane, i.e. are functions only of time. 
The possible contribution of a non zero $\lambda_{brane}$ (defined in (\ref{lagofbra})) will be assumed in the following to be included in $\rho_b$ and $p_b$.
We also 
assume that there is no flow of matter along the fifth dimension, which gets translated to 
\begin{equation}
\tilde{T}_{05}=0. \label{noflow}
\end{equation}
The non vanishing component of $\tilde{U}$ are straightforwardly given by 
\begin{eqnarray} \label{uoo}
\tilde{U}_{00} &=& -\frac{3\delta(y)}{\mu^2 b}\left\{\frac{\dot{a}^2}{a^2} + k\frac{n^2}{a^2}\right\}, \\ \label{uij}
\tilde{U}_{ij} &=& -\frac{\delta(y)}{\mu^2 b} \left\{\frac{a^2}{n^2}\gamma_{ij} 
\left(-\frac{\dot{a}^2}{a^2} + 2 \frac{\dot{a}}{a}\frac{\dot{n}}{n} - 2 \frac{\dda}{a} \right) - k \gamma_{ij} \right\},
\end{eqnarray}
where a dot stands for a derivative with respect to $\tau$.
Our aim is now to obtain the equations governing the cosmological evolution of the induced metric on the brane.

Let us first remind the results obtained by 
Bin{\'e}truy {\it et al.} \cite{bdel} which we will use in the following.    
For the particular setting described above, one can obtain a first integral of the Einstein's equations 
(\ref{einstein}) in the bulk which can be written  
\beq
 \frac{\left(a^\prime  a\right)^2}{b^2} 
-\frac{\left(\dot{a}  a\right)^2}{n^2}-ka^2
+{\K52 \over 6} a^4\rho_B+\C=0 \label{firstint},
\eeq
where $\cal{C}$ is a constant of integration. 
In the above equation, a prime stands for a derivative with respect to
 $y$. 
 One can show that any set
of functions $\{a,b,n\}$ satisfying  (\ref{firstint}) together with 
$\tilde{G}_{05}=0$ (which is here mandatory from (\ref{noflow}))
 will be solution of whole the Einstein's equations in the bulk.
The brane is then taken into account  
by using Israel's junction conditions \cite{Israel66}. These junction conditions
relate the jump across the brane 
of the brane extrinsic curvature to 
the delta functions sources on the right hand side of  Einstein's equations (\ref{einstein}). 
The extrinsic curvature tensor of a given 
$y=\mbox{constant}$ surfaces in the background metric (\ref{cosmback}) is given by 
\beq \label{intcurv}
K^A_B = \mbox{diag} \left(\frac{n^\prime}{nb},\frac{a^\prime}{ab}\delta^i_j,0\right).
\eeq
We find here for the junction conditions (see e.g. \cite{Binetruy:2000ut,bdel} for more details)
\begin{eqnarray}
\frac{\jl a^\prime \jr}{a_0 b_0}&=&-\frac{\K52}{3}\rho_b +
 \frac{\K52}{\mu^2 n_0^2} \left\{\frac{\da_0^2}{a_0^2} +
k\frac{n_0^2}{a_0^2}\right\} , \label{aarho}\\
\frac{\jl n^\prime \jr}{n_0 b_0}&=&\frac{\K52}{3}  \left( 3p_b + 2\rho_b \right)
+ \frac{\K52}{\mu^2n_0^2} \left\{-\frac{\da_0^2}{a_0^2} -2 \frac{\da_0 \dn_0 }{a_0n_0}
 + 2 \frac{\dda_0}{a_0} - k \frac{n_0^2}{a_0^2} \right\}
\label{nnrho}
\end{eqnarray}
where the subscript $0$ for $a,b,n$ means that these functions are 
taken in $y=0$, and $\jl Q \jr = Q(0^+)-Q(0^-)$ denotes the jump
 of the function $Q$ across $y = 0$. We can 
 compare (\ref{aarho}) and (\ref{nnrho}) with the similar equations obtained when 
 we consider  the model given only by action (\ref{action}). These equations can be 
 recovered from (\ref{aarho}) and (\ref{nnrho}) by letting $\mu$ go to infinity. 
  One sees then that the brane intrinsic curvature term (\ref{branac})  acts 
 as a source for the Einstein's equations which depends explicitely on the induced metric 
 time derivatives as well as on its
 spatial curvature. It is also apparent from (\ref{sourcier}), (\ref{uoo}) and (\ref{uij}). 
 For a given induced metric parametrized  by $a_0$, $n_0$ and $k$, the intrinsic curvature term (\ref{branac}) acts as a
`cosmic fluid'\footnote{The 4D Bianchi indentities $\nabla^c G_{cd}=0$  
ensure that the energy-momentum of this `fluid' is conserved.} of density $\rho_{curv}$ and pressure $p_{curv}$ given by  
\begin{eqnarray} \label{rhovurv}
\rho_{curv}&=& - \frac{3}{\mu^2 n_0^2}\left\{\frac{\da_0^2}{a_0^2}+ k \frac{n_0^2}{a_0^2}\right\}
 \label{rhocurv} \\ \label{pcurv}
p_{curv} &=& 
\frac{1}{\mu^2 n_0^2} \left\{
\frac{\dot{a}_0^2}{a_0^2}  - 2 \frac{\da_0 \dn_0}{a_0n_0} + 2 \frac{\dda_0}{a_0}  + k \frac{n_0^2}{a_0^2}  \right\}
\end{eqnarray}
One note that the energy density of this `fluid' is always negative whenever $k=0$ or $k=1$.

Assuming the symmetry $y \leftrightarrow -y$ for simplicity,
 the junction condition 
(\ref{aarho}) can be used to compute $a^\prime$ on the two sides of the brane. We have in this  case 
$\jl a^\prime \jr = 2 a^\prime(0^+)$. 
By continuity when $y \rightarrow 0$,  (\ref{firstint})  yields
the  generalized (first) Friedmann equation:  
\beq
\epsilon \sqrt{{H^2}-{ \K52 \over 6}\rho_B - {\C\over a_0^4}
+  {k \over a_0^2}} =   \frac{\K52}{2\mu^2} \left(H^2+\frac{k}{a_0^2}\right)
 - \frac{\K52}{6}\rho_b ,
 \label{friedfried}
\eeq
where the Hubble parameter $H$ is defined here by 
\beq
H = \frac{\da_0}{a_0n_0},
\eeq
and $\epsilon = \pm 1$ is the sign of  $[a^\prime]$ (see equation
(\ref{aarho})). 
A similar equation has  been obtained in \cite{Shtanov:2000vr}
 for an $AdS$-Schwarzschild
background (in which case $\rho_B$ is negative, see also \cite{Collins:2000yb,Nojiri:2000gv}) but only
 the case $\epsilon = -1$ was discussed there. 
As we will see more explicitely in section (\ref{branebed}) the two different possible 
$\epsilon$ correspond 
to two different embeddings of the brane into the bulk space-time (see also \cite{Gibbons:1993in,Bowcock:2000cq}). 
Before specializing to the case where $\rho_B$ is set to zero, we further note that one can derive for the brane 
matter a usual conservation equation. 
If we  plug into the $(0,5)$ component of the Einstein's equations the jump conditions (\ref{aarho}) and (\ref{nnrho}) 
we obtain, as when no brane intrinsic curvature (\ref{branac}) is included, the conservation equation
\begin{equation}
 \dot{\rho_b} + 3(p_b+\rho_b) \frac{\dot{a_0}}{a_0}  =0.
\label{energy}
\end{equation}
Equation (\ref{energy}) and (\ref{friedfried}), together with the brane matter equation of state, 
are then sufficient to 
derive the cosmological evolution of the brane metric.
Last, we recall the brane-Friedmann equation obtained \cite{bdel,Shiromizu:2000wj,Flanagan:2000cu} 
when no curvature term (\ref{branac}) is included. It can easily be deduced from 
(\ref{friedfried}) by letting $\mu$ go to infinity and reads (we write it here  for a zero bulk 
cosmological constant and a zero Schwarzschild mass parameter ${\cal C}$)
\beq \label{555}
H^2 +\frac{k}{a_0^2}=\frac{\kappa^4}{36}\rho_b^2.
\eeq
We will refer to a regime where such an equation is (approximately) 
valid as a {\it fully 5D regime}.

\section{Discussion}
We want now to discuss the solutions to the Friedmann-like equation (\ref{friedfried}) together with (\ref{energy})
when the bulk cosmological constant $\rho_B$ vanishes. 
We recall that one can interpret the constant ${\cal C}$ 
appearing in (\ref{friedfried}) as coming from the 5 dimensional bulk Weyl tensor 
\cite{Shiromizu:2000wj,Mukohyama:2000wi}, since we are mainly interested here in 
Minkowskian bulk (for which the Weyl tensor
 vanishes) we will also set ${\cal C}$ to zero in the following discussion.

\subsection {Recovery of standard cosmology}
 Equation (\ref{friedfried}) can straightforwardly be rewritten
\begin{equation} \label{refrite}
\frac{\mu^2}{3}\rho_b = H^2 + \frac{k}{a_0^2} - 2 \epsilon \frac{\mu^2}{\kappa^2}\sqrt{H^2 + \frac{k}{a_0^2}}.
\end{equation} 
It is now apparent on (\ref{refrite}) that the standard cosmology, namely the usual 4D Friedmann equation:
\beq \label{usualfried}
\frac{8 \pi G_{(4)}}{3}\rho_b = H^2 + \frac{k}{a_0^2},
\eeq 
is recovered whenever the last term on the right hand side
of (\ref{refrite}) is subdominant with respect to the first two terms, namely when
\begin{equation}
\sqrt{H^2 + \frac{k}{a_0^2}} \gg 2  \frac{\mu^2}{\kappa^2},
\end{equation}
or in terms of the Hubble radius $H^{-1}$, when (neglecting the spatial curvature term)
\begin{equation} \label{borneopp}
H^{-1} \ll \frac{M^2_{(4)}}{2 M^3_{(5)}}.
\end{equation}
This matches the scale $r_0$ (\ref{crosso}) found by Dvali {\it et al.}   
setting the crossover between the 4D gravity and the 5D gravity regimes.
We will however show in the next subsection
 that for larger Hubble radii, one does not necessarily enter into a fully 5D regime
(\ref{555}). 
We want here to emphasize that, even if the evolution of the universe when (\ref{borneopp}) is verified 
is close to the standard one,
 there is
a mismatch between the `Newton's constant' $G_{(4)}$ derived from the Friedmann equation (\ref{refrite}),
and simply given in terms of $\mu^2$ by the
usual relation  (\ref{geengri}),
and $G_N$ as defined in (\ref{newton}). 
This mismatch can be tracebacked to the presence of the extra scalar degree of freedom. 
We will come back to this question in the last section.

\subsection{Late time cosmology}
For Hubble radius for which (\ref{borneopp}) is not satisfied,
  there are two distinct behaviours depending on $\epsilon$. 
 Equation (\ref{refrite}) can indeed be rewritten (assuming $\rho_b \geq 0$)
\beq
\sqrt{H^2+\frac{k}{a_0^2}} = \epsilon \frac{\mu^2}{\kappa^2} + \sqrt{\frac{\mu^2}{3}\rho_b +  \frac{\mu^4}{\kappa^4}}
\eeq

Let us start by examining the case where $\epsilon = -1$. It is easy to show,
integrating the above equation, that for $k=0$ or $k=-1$, and assuming for simplicity\footnote{the following discussion is
also valid in a more general case where the brane energy density is the sum of that of different kinds of cosmic `fluids':
matter, radiation, cosmological constant, as long as this energy density reaches a 
regime where it is small in comparison with
$\mu^2/\kappa^4$.}
some usual equation of state for the brane matter
of the form 
\beq \label{eqstate}
p_b = w \rho_b \quad \quad \mbox{(with $w \geq -1$),} 
\eeq
that $a_{0}$ diverges for late time, so that the density of any matter (with $w > -1$) goes to zero for late time,
and reaches a regime where it is
small in comparison with $\mu^2/\kappa^4$ (which is equivalent to saying that $\sqrt{H^2+k/a_0^2} \ll r_0^{-1}$)
\footnote{when $k=1$, it is possible that the universe turns over before reaching the regime where 
$\rho_b \ll \mu^2/\kappa^4$.}.
 One can then expand (\ref{refrite}) to obtain 
\beq
\sqrt{H^2+\frac{k}{a_0^2}} = \frac{1}{6}\kappa^2\rho_b, 
\eeq
which is the fully 5D regime (\ref{555}).
One has thus a transition from a 4D regime to a 5D regime.

If  $\epsilon = 1$, however,  
 $\sqrt{H^2 + \frac{k}{a_0^2}}$ is always larger than $H_{self}$ given by
\begin{equation}
H_{self} = \frac{2\mu^2}{\kappa^2}.
\end{equation}
And the expansion will never enter into a  fully 5D regime.
If we start from initial conditions verifying (\ref{borneopp}) (and again if $k=0$ or $k=-1$), it is easy to show, since 
 $H$ is bounded from below by $H_{self}$ that the scale factor $a_0$ goes to infinity at large time, which implies
  again that
 the brane matter energy density goes to zero for any matter 
 verifying 
(\ref{eqstate}) with $w > -1$. In this quite generic case, one 
 has  a transition between a usual 4D FLRW cosmology governed by equation ({\ref{usualfried}) and an inflationary 
solution where $H$ is (approximately) constant and given by $H_{self}$. The $\epsilon =1$ case is thus able to give a 
quintessence-like scenario where a phase of matter or radiation
dominated cosmology is followed by a late phase of accelerated expansion.

Shtanov \cite{Shtanov:2000vr} (see also \cite{Nojiri:2000gv}), already noticed that setting $\rho_b$ to zero in the
brane Friedmann equations, one obtains,
 aside from the trivial $\dot{a}=0$ solution, a de Sitter like expansion, the self-inflationary solution quoted above. 
This last solution can easily be understood if we look at the `cosmic fluid' defined by
(\ref{rhocurv}) and (\ref{pcurv}) in the case where we set $H$ to a constant value. 
We have then, neglecting the spatial 3D curvature 
\beq
\rho_{curv}= -p_{curv}= -\frac{3H^2}{\mu^2} = \mbox{constant}, 
\eeq
so that the intrinsic curvature term (\ref{branac}) acts as a negative cosmological constant on the brane. Since the 
sources on the brane 
enter quadratically the Friedmann-like equation (\ref{friedfried}) it can however lead to a de Sitter expansion governed by
the brane Friedmann equation (\ref{555})
\begin{equation}
H^2 = \frac{\kappa^4}{36} \rho^2_{curv},
\end{equation} 
which leads in turn to a given value for $H$, namely $H_{self}$. 
For this particular $H$, the induced metric is thus identical to the one, one would obtain from a brane without a self
curvature term (\ref{branac}) but with a positive brane tension $\rho_b$ given by 
\begin{equation}
\rho_b = - \rho_{curv}.
\end{equation}
This last solution will be called in the following the TI (tension-inflationary) solution,
 whereas the self-inflationary solution will be
referred to as the SI solution. 
Let us further stress here that 
the sign of $-\epsilon$ indeed measures the sign of the 
effective energy density on the brane as seen from the 5D bulk theory.
This effective energy density can be defined  from (\ref{aarho}) and (\ref{rhocurv}) as
 the sum of the brane matter
energy density $\rho_b$ and the curvature energy density $\rho_{curv}$. When $\epsilon$ is equal to 1, 
this   
effective energy density  is  negative (even if $\rho_b$ is positive), so that it is possible that the $\epsilon =1$ 
scenario
can be plagued by some instability associated with a violation of a positive energy condition.

\subsection{Brane embedding in Minkowski space-time} \label{branebed}
We want here to discuss briefly the embedding of the cosmological solutions into 
Minkowski space-time. 
We follow again \cite{bdel} to obtain the bulk metric under the assumption 
that $b$ does not depend on time (so that with an
appropriate redefinition of $y$, one can assume that $b=1$).
Having again (\ref{noflow}), the $(0,5)$ component of Einstein's equations lead to 
\beq
{\dot a\over n}=\alpha(\tau), \label{alpha}
\eeq
where $\alpha$ is a function that depends only on time (and not on $y$).
By  a suitable change of time one can choose $n_0=1$ so that $\alpha$ is simply given by $\dot{a}_0$.
On the other hand, The $(0,0)$ component of Einstein's 
equations can be straightforwardly integrated in the bulk to lead to \footnote{when $\rho_B = 0$ and assuming 
again $y \leftrightarrow -y$ symmetry. We refer the interested reader to \cite{bdel} for more details.}
\beq
a^2=\left(\dot{a}_0^2+k\right) y^2+a_0 [a^\prime]|y|+a_0^2.
\eeq
which can in turn be expressed in terms of the induced metric 
and the energy-momentum tensor on the brane through the jump condition (\ref{aarho}).
The scale factor $a$ is thus given by 
\beq
a=a_0 \left\{ 1 + |y|\left(-\frac{\K52}{3}\rho_b +
 \frac{\K52}{\mu^2} \left(H^2 +
\frac{k}{a_0^2}\right)  \right) + y^2 \left(H^2 + \frac{k}{a_0^2}\right) \right\}^{1/2} \label{aaa},
\eeq
 and $n$ can be obtained by (\ref{alpha}). 
The unknown functions $a_0$ and $\rho_b$ in (\ref{aaa}) (as well as in the expression of $n$) should then be 
computed through
equations (\ref{friedfried}) and (\ref{energy}) as well as by the use of the equation 
of state for the brane cosmic fluid.
Considering again the ${\cal C}=0$ case, and using (\ref{friedfried}), the bulk metric is given by
\begin{eqnarray} \label{bulkmet}
a_\epsilon &=& a_0 + \epsilon |y| \left(\dot{a}_0^2+k \right)^{1/2},  \\
n_\epsilon &=& 1 + \epsilon |y| \ddot{a}_0\left(\dot{a}_0^2+k \right)^{-1/2}, \nonumber  \\
b_\epsilon  &=& 1. \nonumber
\end{eqnarray}
This space-time  can be shown  to be a slice of Minkowski space-time. 
Indeed in the particular case of $k=0$ and $\epsilon = -1$ , Deruelle and Dolezel \cite{Deruelle:2000ge}
obtained an explicit change of coordinate $Y^A = Y^A(X^A)$ to go to the canonical 5D Minkowskian metric  
\begin{equation} \label{canon}
ds^2 = -(dY^0)^2 + (dY^1)^2 + (dY^2)^2 + (dY^3)^2 + (dY^5)^2.
\end{equation}
One can check that this change of coordinate is also valid for $\epsilon = +1$, it is given by 
\begin{eqnarray}
Y^0 &=& a_\epsilon \left( \frac{r^2}{4}+1-\frac{1}{4\dot{a}_0^2} \right) -\frac{1}{2}
 \int \frac{a_0^2}{\dot{a}_0^3} \partial_\tau
\left(\frac{\dot{a}_0}{a_0}\right) d\tau, \label{changun}\\
Y^i &=&  a_\epsilon x^i,\nonumber \nonumber \\ \nonumber 
Y^5 &=& a_\epsilon \left( \frac{r^2}{4}-1-\frac{1}{4\dot{a}_0^2} \right) -\frac{1}{2} 
\int \frac{a_0^2}{\dot{a}_0^3} \partial_\tau
\left(\frac{\dot{a}_0}{a_0}\right) d\tau ,
\end{eqnarray}
where $r^2= x^i x^j \eta_{ij}$, and $\eta_{ij}$ is here a flat euclidian 3D metric.
A related work is the one of Mukhoyama {\it et al.} \cite{Mukohyama:2000wi} 
which gives the change of coordinates 
from a static Schwarzschild-$AdS$ bulk metric \cite{Kraus:1999it} 
to a gaussian normal system as we used in this work (see also \cite{Bowcock:2000cq}). 
For the metric  (\ref{bulkmet}) with $k \neq 0$, one can
 find as well a change of coordinate $Y^A = Y^A(X^A)$ leading to the canonical 
 Minkowski metric (\ref{canon}).
 It can be defined for $k= 1$ by 
\begin{eqnarray}
Y^0 &=& a_\epsilon \tilde{y} + \tilde{z}, \label{changdeux} \\
Y^i &=& a_\epsilon \tilde{Y}^i,  \nonumber \\
Y^5 &=& a_\epsilon \tilde{Y}^5,  \nonumber
\end{eqnarray}
where $\tilde{Y}^i$ and $\tilde{Y}^5$ are functions of the $x^i$ only and verify 
\beq
(\tilde{Y}^5)^2 + \sum_{i=1,3} (\tilde{Y}^i)^2 = 1,
\eeq
which defines a 3D $k=1$ maximally symetric  space, and $\tilde{y}$ and $\tilde{z}$ are given by
\begin{eqnarray} \label{tildy}
\tilde{y} &=& \frac{\dot{a}_0}{\sqrt{\dot{a}^2_0 +k}},  \\
\tilde{z} &=& k \int \frac{1}{\dot{a}_0} \partial_\tau \left(
\frac{a_0}{\sqrt{\dot{a}_0^2+k}} \right) d\tau. \label{tildz}
\end{eqnarray}
Similarly for $k=-1$, we find 
\begin{eqnarray}
Y^0 &=& a_\epsilon \tilde{Y}^0, \label{changtrois} \\ 
Y^i &=& a_\epsilon \tilde{Y}^i,  \nonumber \\
Y^5 &=& a_\epsilon \tilde{y} + \tilde{z}, \nonumber
\end{eqnarray}
where $\tilde{Y}^i$ and $\tilde{Y}^0$ are function of the $x^i$ only and verify 
\beq
-(\tilde{Y}^0)^2 + \sum_{i=1,3} (\tilde{Y}^i)^2 = -1,
\eeq
which defines a 3D $k=-1$ maximally symetric space,  
$\tilde{y}$ and $\tilde{z}$ are still given by (\ref{tildy}) and (\ref{tildz}).
The change of coordinates (\ref{changun}), (\ref{changdeux}) and (\ref{changtrois}) are of the form
\beq
Y^A = a_\epsilon \left(\tilde{Y}^A + \tilde{y}^A\right) + \tilde{z}^A, 
\eeq
with $\tilde{Y}^A$ being a function of the $x^i$ only, $\tilde{y}^A$ and $\tilde{z}^A$ 
are functions of time and do not depend on $\epsilon$. One can then invert these change of coordinate to obtain
\beq \label{epi}
\epsilon |y| = f(Y^A), 
\eeq
with $ f(Y^A)$ a function of the $Y^A$ independent of $\epsilon$.
The brane is then the hypersurface ${\cal H}$ (or a slice of it, see below) in Minkowski space-time defined by $f(Y^A) = 0$.
We see clearly on (\ref{epi}) that for a given $\epsilon$ (see \cite{Gibbons:1993in,Deruelle:2000ge} )
 the coordinate system $y,\tau,x^i$ provides a double covering of only  
one side of ${\cal H}$.
The two copies of the side of ${\cal H}$ which is kept are respectively given by the $y>0$ and $y<0$ half-spaces 
(in the coordinates $X^A$) and are glued
together along the brane; this provides in turn a jump of  the extrinsic curvature tensor across the brane.
Moreover, for a given function ${a}_0$, the $\epsilon = \pm 1$ choice 
corresponds to keeping one side or the other
 of ${\cal H}$.
It is also clear from (\ref{epi}), as well as from the definition of $\epsilon$,
that the jump in the extrinsic curvature for the $\epsilon = \pm 1$ solutions and a given function $a_0$ 
are opposite to each other.

One can check for example that when
\beq \label{de Sitt}
\frac{\dot{a}^2_0}{a^2_0} + \frac{k}{a_0^2}=H^2_0,
\eeq 
with $H_0$ being a constant and $k=1$, the flat coordinates $Y^A$ verify   
\beq \label{hyperbol}
(Y^5)^2 + (Y^3)^2+ (Y^2)^2 +(Y^1)^2 -(Y^0)^2 = \frac{\left(1 + \epsilon H_0 |y|\right)^2}{H_0^2}.
\eeq
We recover in this case the standard embedding of de 
Sitter space-time (here, the brane) into a 5D Minkowski space-time as a hyperboloid\footnote{one can indeed verify from 
(\ref{changdeux}) that the brane covers the whole hyperboloid.} ${\cal H}_0$
defined by setting $y$ to zero in equation (\ref{hyperbol}).
For the inflationary solution ({\ref{de Sitt}) with $k=0$ (and still $H_0$ a constant)
 one find easily that (\ref{hyperbol}) is still verified with the further restriction 
that $Y^0(y=0,\tau)-Y^5(y=0,\tau) >0$ (as can be seen from (\ref{changun})). We recover the 
well known result (see e.g. \cite{hawk}) that this inflationary 
solution only covers half of de Sitter space-time. For $\epsilon =-1$ (and still
$k=0$), this is also 
in agreement with 
the interpretation of the $Z_2$ symmetric domain-wall space-time of refs.
\cite{domainwall} as the interior of ${\cal H}_0$
\cite{Gibbons:1993in}. This latter case  
 corresponds with the TI solution, while the SI solution 
 covers
the exterior of ${\cal H}_0$.

Eventually, we note that the dependence of the metric (\ref{bulkmet}) on $a_0$ is the same as when no brane intrinsic
curvature term (\ref{branac}) is included. So that the above discussion also applies in this case.
 The only difference between
the two cases (with and without a brane intrinsic
curvature term (\ref{branac})) is the dynamics of $a_0$, that is to say the brane trajectory into the bulk space-time (see
\cite{Bowcock:2000cq} for a very complete study of this question).

\section{Conclusions}
Let us first summarize some of our main results.
We have studied the cosmology of a $Z_2$ symmetrical 3-brane embedded in a 5D Minkowskian
space-time, when an intrinsic curvature term is added on the brane. We have shown in particular
that the usual cosmology is recovered for Hubble radii smaller than the crossover scale between 4D and 5D
gravity found by Dvali {\it et al.} \cite{Dvali:2000hr} and given by $r_0= M_{(4)}^2 / 2 M^3_{(5)}$.
 If we consider a matter content (radiation, matter,...)  such that the energy density is 
decreasing as the scale
factor increases (and for $k =0$ or  $k=-1$), 
then one has two possibilities depending on the initial conditions. The universe
 either evolves
toward a fully 5D regime where the relation between the Hubble parameter and the energy density is
linear, or it evolves toward a self-inflationary solution, where the inflation is sourced by the scalar
curvature term on the brane. 
We have also shown  that the Newton's constant which enters the Friedmann-like equation
during the early standard like evolution of the universe differs from the one defined from the measurement of the 
gravitational force
between static point sources.

Although this work was mainly intended in an effort to better understand the models with brane and
intrinsic curvature term on the brane, and could shed some light on the most interesting cases where there is more than one 
 extra dimension, one could also look at it in a more phenomenological
perspective.  If one wishes to have an acceptable model 
from a phenomenological point of view, one should first take care of the 
presence of the extra scalar degree of freedom in the graviton propagator with respect to standard 4D gravity.
As far as cosmology is concerned, 
this extra degree of freedom is responsible for the difference between the 
`Cavendish' Newton's constant $G_{N}$ 
and the `cosmological' Newton's constant
$G_{(4)}$. As a consequence of that, if we set $G_{N}$ to its measured value, 
$G_{(4)}$ will be smaller by a factor $3/4$ than the usual Newton's constant entering into the usual Friedmann equations.
This is for exemple responsible for a change in the rate of expansion during
nucleosynthesis and may be marginally compatible we the actual bounds. However, this extra scalar 
degree of freedom can not be compatible with relativistic tests of General Relativity \cite{Dvali:2000hr},
 and one should first of all be able to cancel its effect. Although some proposal exist \cite{Dvali:2000hr}, like adding an extra vector field in
 the model, this seems very hard to do in a realistic way.
Another crucial question, as acknowledged by Dvali {\it et al.}, is whether the crossover scale $r_0$ 
can be made large enough\footnote{Indeed the present work enables to put constraints on $r_0$ from cosmology
 in order for the transition to the fully 5D regime or to the self-inflationary solution to happen at late enough time.}.
Setting $M_{(5)}$ to $TeV$ (and knowing $M_{(4)}$) gives a scale \cite{Dvali:2000hr} that is much
 too low,  so that one
needs a very small 5D Planck scale which seems very difficult to cook.
After this pessimistic account, we would like to underline a possible virtue of the hereabove described solutions, 
 namely the fact that the $\epsilon =1$ scenario  leads  to
an interesting alternative to produce a phase of late accelerated expansion 
as indicated by the SN data \cite{supernovae}. We note moreover, that this last scenario `explains'
in a natural way the
right order of magnitude for the crossover scale (in terms of the Hubble radius) between matter dominated cosmology and late
accelerated expansion: this scale is set by $r_0$ which is also the crossover scale between 4D and 5D gravity and has
thus to be of cosmological size (see e.g. \cite{Dvali:2000hr,Binetruy:2000xv}). It would be interested to know if
such features  persist in a more acceptable model with one extra dimension
or for more than one transverse extra dimension.

\section*{Acknowledgments} 
We wish to thank B.Bajc, G.Dvali, A.Lue and M.Porrati for stimulating and 
enlightening discussions and comments on a draft of this paper.
This work is sponsored in part by NSF Award PHY 9803174,
 and by  David and Lucile Packard Foundation
Fellowship 99-1462.

\end{document}